# Electron transport in multi-terminal molecular device

Kamal K. Saha,[1] Wenchang Lu,[1,2] J. Bernholc,[1,2] and Vincent Meunier[1,3,*]

[1]*Computer Science and Mathematics Division, Oak Ridge National Laboratory, Oak Ridge, Tennessee 37831-6367, USA*

[2]*Center for High Performance Simulation and Department of Physics, North Carolina State University, Raleigh, North Carolina 27695-7518, USA*

[3]*Center for Nanophase Materials Science, Oak Ridge National Laboratory, Oak Ridge, Tennessee 37831-6367, USA*

**Abstract.** The electron transport properties of a four-terminal molecular device are computed within the framework of density functional theory and non-equilibrium Keldysh theory. The additional two terminals lead to new properties, including a pronounced negative differential resistance not present in a two-terminal setup, and a pseudo-gating effect. In general, quantum interference between the four terminals and the central molecule leads to a complex non-linear behavior, dependent on the alignment of individual molecular states under bias and their coupling to the leads.

[*]**corresponding author:** meunierv@ornl.gov

In recent years, numerous papers have been published to investigate the connection between the microscopic characteristics of an electronic system, such as atomic configuration and electronic structure, and the electron transport properties, such as electrical current and conductance.[1-4] These results have substantially improved the general understanding of the I-V characteristic of nanojunctions. While it is still difficult to fully determine the atomic geometries of measured junctions and to establish a one-to-one correspondence between the observed and calculated results, the progress has been rapid in both experimental and theoretical capabilities. To date, most or all existing *ab initio* approaches to electron transport deal only with two-terminal systems. It is however of general interest to develop robust computational schemes that can routinely and reliably account for the transport mechanism in multi-terminal molecular junctions, as these are critically important in both fundamental studies and applications. Even in classical measurements of key electrical properties, four-terminal techniques are vastly more accurate and reliable than the two-terminal ones. Although such measurements are very difficult in nanoscale systems, four-probe scanning tunneling microscopy (STM) enables a rather complete characterization of nanoscale conductance and of various potential building blocks for molecular and nanoscale electronics.[5] Furthermore, multi-terminal structures are critical for applications. For instance, it is impossible to achieve amplification using only two terminal structures. [6] This is because all electronic circuits have some loss and therefore require multistage amplification. While simple logic operations could in principle be performed without amplification, modern computing operations consist of many such steps, which inevitably result in the need for amplification. This need is even greater in sensors and antennas, which must greatly amplify very weak signals to ensure detection and communication. Despite the importance of multi-terminal devices, multi-terminal aspects at the nanoscale level have hardly been investigated either experimentally or theoretically, except in the "field-effect-transistor" geometry, in which a gating potential is applied through a well-insulated macroscopic gate, without any possibility of leakage current though the gate electrode.[6] While this geometry is very useful in classical devices, the effects of leakage current in molecular-scale devices can be substantial. In addition, quantum effects enter at this scale, potentially resulting in novel and unforeseen phenomena.

In this Letter, we describe the results of a state-of-the-art *ab initio* study of a four-terminal device consisting of an organic molecule connected to four gold nanowires. We use a newly developed multi-terminal formalism,[7] and employ self-consistent density functional theory in the context on a non-equilibrium Green function method. We find that the presence of additional terminals leads to new effects, including the emergence of a strong negative differential resistance (NDR), not present in the corresponding two-terminal device, and unexpected quantum interference patterns that can be explored in devices.

The molecular system investigated here is represented in the inset of Fig. 1. It consists of a [9, 10-Bis((2'-para-mercaptophenyl)-ethinyl)-anthracene] molecule connected via thiol bridges to four gold nanowires. While this system is challenging to realize experimentally with current nano-manipulation techniques, similar geometries can be achieved using a four-probe STM for molecules deposited on an insulating substrate. In a two-terminal geometry, two main factors govern the I-V characteristics: the position and properties of orbitals originating from the free molecule (we call them pseudo-molecular orbitals), and details of the contacts between the molecule and the electrodes. While these factors are also important in multi-terminal setups, the intricacies of mutual couplings between the individual probes give rise to even more complex I-V characteristics. The molecule shown in Fig. 1 has already been investigated in a two-terminal configuration. Reichert *et al.* [8] reported an experimental investigation of this and a related molecule in a break junction geometry. While the precise atomic structures of the contacts were not determined, the authors presented strong arguments that they have measured single molecule I-V characteristics. Heurich *et al.* [9] performed DFT-tight-binding calculations for the same system, obtaining results in qualitative agreement with experiment. In both experiment and theory, the I-V curve is symmetric about the zero voltage line and smooth, with no negative differential resistance (NDR) effects.

The present calculations are carried out using a recently developed multi-terminal formalism that allows for self-consistent DFT calculations of the I-V characteristics of complex nanojunctions.[7] All quantities are computed at the density functional theory (DFT) level with full self-consistency (SC) under applied bias, using ultrasoft pseudopotentials [10] and Perdew-Burke-Ernzerhof exchange-correlation functional.[11] We follow a four-step approach: First, sets of atom-centered localized orbitals are optimized variationally for the leads and the central molecule separately, and their electronic structures are obtained at the same time.[12] We use 10, 6, and 4 orbitals per Au, C, and H atoms, respectively, all with radii of 4.3 Å. Each of the four electrodes is simulated by an Au(111) nanowire,[13] built from (111) subunits. The periodic subunit is made up of three layers containing 7, 6, and 6 atoms, respectively. The extended scattering region $S$ includes the molecule and 8 layers of Au in each lead. The inclusion of the 8 "extended buffer" layers ensures a proper treatment of the molecule-lead coupling,[14] as well as proper screening of the potential drop in the leads. Second, the conductor's Green function, coupling matrices and self-energies are expanded within this basis set, and the charge density and potentials are calculated self-consistently for the combined system at zero bias. During the self-consistency process, the Hartree potential is obtained by solving the Poisson's equation in the $S$ region with boundary conditions that match the electrostatic potentials of all the leads. Third, after the charge density for the combined system is converged at zero bias, a realistic bias potential is applied.[15] The non-equilibrium Green function (NEGF), charge density, and potentials of the four-terminal junction are computed through a self-consistent process [7] using the NEGF formalism.[15, 16] Finally, the current from the $i$ through the molecular barrier to lead $j$, driven by the bias $V = \mu_i - \mu_j$, is obtained from

$$I_{ij}(V) = \int_{-\infty}^{\infty} T_{ij}(\varepsilon, V)[f(\varepsilon - \mu_i) - f(\varepsilon - \mu_j)]d\varepsilon,$$

where $f$ is the Fermi function and $T_{ij}$ is the energy- and voltage-dependent transmission spectrum from the $i$-th to $j$-th lead, which also self-consistently depends on the voltages applied to the other leads through the Hartree potential acting on the electrons.

In the DFT-optimized geometry presented in Fig. 1, two ends of the molecule are covalently connected to Au leads via thiol bridges. The other two ends are connected to Au leads via thiophene groups [17]. For a two-terminal setup without leads 3 and 4, our calculated *I-V* curve (Fig. 1) shows no NDR features, in agreement with experiment. The absolute value of the current is substantially larger than the experimental one.[8] Apart from the likely difference in contact geometries between our calculations and the experiment, it is well-known that DFT underestimates the HOMO-LUMO gap, which also results in an overestimate of the current.[18] Our current is also larger than that computed previously.[9] In the latter calculation, a single Au atom was used to connect the molecule to the Au electrode, and the electrode-molecule coupling was determined by a tight-binding parameter. These two differences can largely explain the difference with our DFT-based calculation using more realistic leads.

In a four-terminal junction, many distributions of bias voltages are possible. In the following, we consider two paradigmatic cases. In the first one, an identical bias potential V/2 is applied at leads 1, 3 and 4, while –V/2 is applied to lead 2 (see the inset in Fig. 1). Using this bias distribution, we find that the $I_{12}$ current (Fig. 1) is asymmetric and presents a pronounced NDR. This finding is in striking contrast with the two-terminal result, where the I-V characteristic is much smoother, with only a weak structure present and no NDR. In this four-terminal geometry the $I_{32}$ current (Fig. 1) is very different from $I_{12}$, showing pronounced asymmetry between the forward and reverse directions, as well as a weak NDR in the forward direction.

The shapes of the different I-V curves in the four-terminal geometry can be explained by analyzing the behavior of individual pseudo-molecular states under bias and the relative strength of their coupling with the four electrodes. The position-dependent density of states (DOS) along the left-to-right direction is shown in Fig. 2. The zero of energy is chosen to be the average of the chemical potentials of lead 1 and lead 2, $(\mu_1 + \mu_2)/2$. The LUMO state, located at 74 meV (dashed-oval) at the equilibrium (no bias applied), follows rigidly the changes in $\mu_1$ and extends all the way from the left to the right electrode. The HOMO state, located at -0.44 eV (dotted-oval) at the equilibrium, is spatially split between the left and right electrodes, each end being pinned to major features of its respective lead. As a result, when parts of the continuum and bias-broadened HOMO reach the bias window, its left and right tails have little overlapping density. This situation occurs for both positive and negative biases, and shows that the LUMO contributes the most to the total current, with some variation due to the details of the bias-dependent coupling with the leads.

To examine this coupling, we plot in Fig. 3 the local DOS's near the lead-molecule interfaces for various bias potentials. The left-to-right I-V curve ($I_{12}$ in Fig.1) can be understood by monitoring the position of the LUMO (red arrows). As the bias increases from 0.0 to 0.8 V, the width of the LUMO slowly increases until 0.2 V, resulting in an essentially linear increase in the current with an increase in the bias window (marked by dotted vertical lines in Fig 3.). At 0.5 V, the coupling with the electrodes suddenly decreases due to the action of the potential applied to the three electrodes at positive bias. The strong influence of the bias also results in a rearrangement of the charge distribution in the molecule. This leads to a uniform electrostatic potential drop from the left to the right lead, and the corresponding electric field shifts the electron density towards the left lead, eventually resulting in a non-linear *I-V* curve. As the bias increases further, the current decreases sharply until the coupling (broadening) increases again, starting at 0.8 V. A similar explanation holds for the negative bias, with slight variations in the positions and spreads of HOMO and LUMO within the bias window, which account for the changes in current amplitudes.

Turning to the bottom-to-right ($I_{32}$) or top-to-right ($I_{34}$) currents, which are equal by symmetry, the current amplitudes for the positive bias are substantially smaller than those for the negative bias. The positive bias induces a possible current-carrying level at the top, bottom, and left sides, but there is no corresponding level at the right collecting electrode, and the current remains small. The situation is different for negative bias, where a current-carrying level exists at the top, bottom, and right electrodes. This level is quite broad, leading to efficient coupling and thus a large current. The apparent asymmetry in the positive and negative biases is thus due to different effects of positive and negative biases on the pseudo-molecular states of the central region.

As stated above, various combinations of bias distributions are possible. Here we focus on a distribution of applied bias potentials that mimics the presence of a pseudo-gate voltage $V_g$, applied to the top (terminal 4) and bottom (terminal 3) electrodes, as shown in Fig. 4(a). The source (terminal 1) and drain (terminal 2) voltages are fixed at +0.1 and -0.1 V respectively. The pseudo-gate voltage is swept from -0.2 to 0.2 V (Fig. 4a). This is not a typical gate (hence the adjective "pseudo"), because the pseudo-gate is not separated from the source-drain channel by an insulating layer and a quite large leakage current is expected, in light of the results shown above. The currents $I_{ij}$, flowing from terminal $i$ to $j$ are shown in Fig. 4b. As the pseudo-gate voltage $V_g$ is increased, the source-to-drain current $I_{12}$ decreases. At equilibrium (no applied bias in any terminal), the LUMO state of the central molecule is just above the Fermi level of the leads. The LUMO state is shifted down by negative drain voltage and negative $V_g$, aligning with the Fermi energy of the source when $V_g = 0.2$ V. This effect results in a large absolute current and the NDR feature (see Fig. 1 for positive bias), as explained above. When $V_g$ is positive, the LUMO state is still shifted down by the negative drain voltage, but this shift is not as large as that with negative $V_g$. The gate-to-drain current $I_{32}$ increases and the source-to-gate current $I_{13}$ decreases as $V_g$ increases from -0.2 to 0.2 V. Fig. 4b also shows that the leakage currents $I_{32}$ and $I_{31}$ are much smaller in absolute values than the source-to-drain current $I_{12}$. By symmetry, $I_{41}$ and $I_{42}$ are identical to $I_{31}$ and $I_{32}$. Fig. 4c depicts the total currents at terminal $i$, $I_i = \sum_{j=1}^{4} I_{ji}$. $I_i > 0$ indicates electrons entering the terminal $i$ and $I_i < 0$ indicates electrons leaving the terminal $i$. The total current $I_1 = I_{21} + I_{31} + I_{41}$, flowing to the source, increases (the absolute value of this current decreases) with increasing gate voltage. This is due to the fact that both $I_{21}$ and $I_{31}$ ($I_{41}$) increase with the increase of the gate voltage. In contrast, $I_{12}$ decreases at the same rate as $I_{32}+I_{42}$ increase. As a result, the total current, $I_2 = I_{12} + I_{32} + I_{42}$, flowing into the drain (Fig. 4c), is not affected by the pseudo-gate voltage. This feature could potentially be used in a future device application. Since the leads 3 and 4 remain at the same pseudo-gate potential $V_g$, there is no current flow between them. At the fixed source-drain voltage of 0.2 eV, the pseudo-gate voltage of 0.02 V results in no current flowing through the pseudo-gate electrodes ($I_3=I_4=I_{13} + I_{23}= 0$). This result provides key information about the electrostatic potential distribution inside the central region, which can be experimentally probed to characterize a device under non-equilibrium conditions.

In summary, we have presented a DFT study of a four-terminal molecular system, highlighting features that are absent in conventional two-terminal setups. The results show counter-intuitive features induced by the quantum-mechanical interplay between the four terminals, including the introduction of a large negative differential resistance that is absent in a two-terminal geometry. The currents between the different terminals are dissimilar and highly non-linear, due to the complex effect of a spatially intricate bias potential interacting with electronic levels of the central region. While the flexibility of a multi-terminal device and its varying responses to biases applied at different terminals remain to be explored and categorized, they open the possibility of novel, multi-functional device structures with nanoscale dimensions.

This research was sponsored in part by the Laboratory Directed Research and Development Program of Oak Ridge National Laboratory (ORNL), managed by UT-Battelle, LLC for the U. S. Department of Energy under Contract No. De-AC05-00OR22725. Work at NCSU was partially funded by ONR N000140610173.

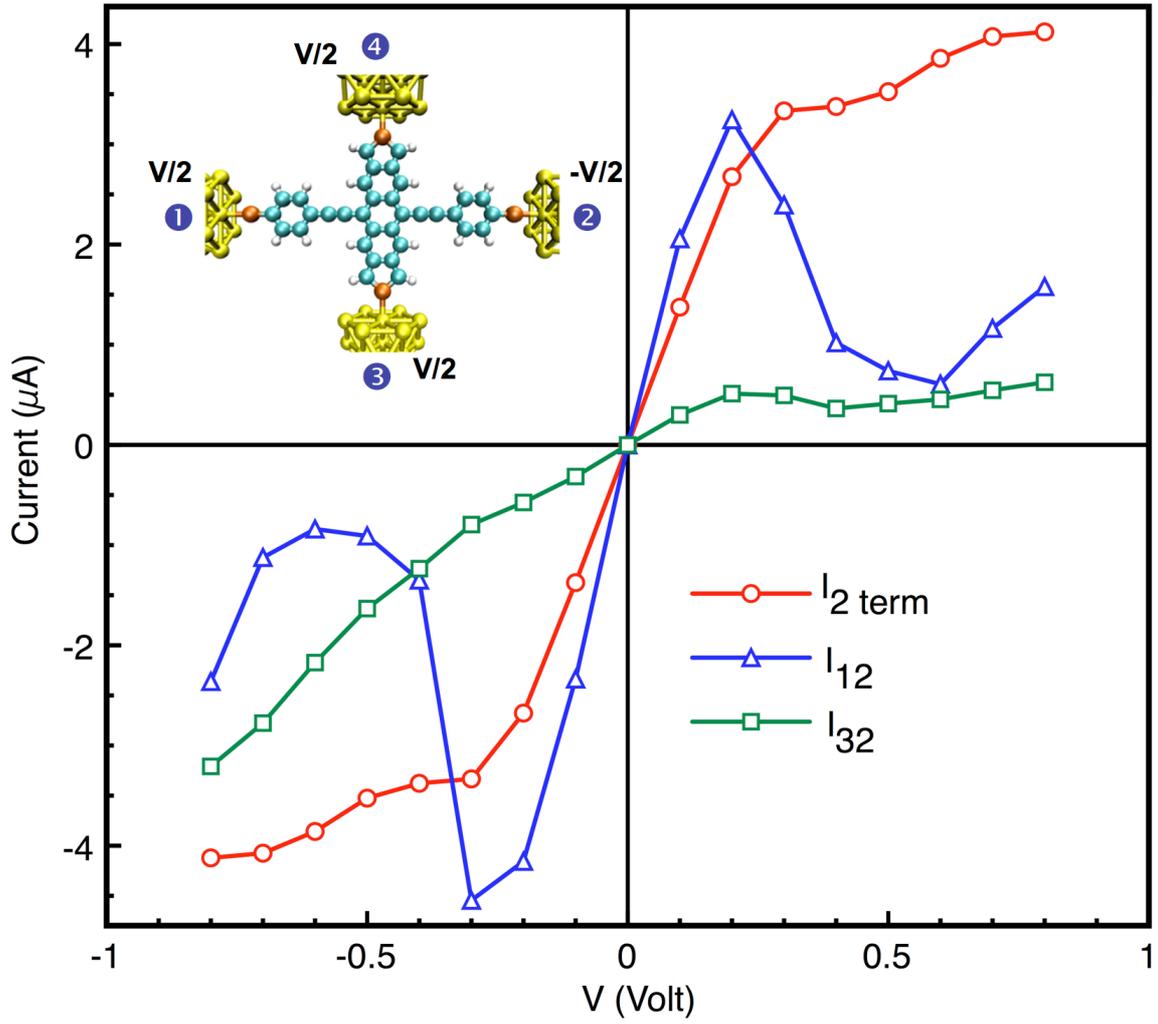

FIG. 1: (Color online). I-V curves of the four-terminal molecular device. For comparison, the I-V curve ($I_2$) of the corresponding two-terminal 1-2 setup is also shown. In the inset, the bias geometry of the four-terminal system is schematically displayed. Au, S, C and H atoms are shown in yellow, red, cyan and gray, respectively. The semi-infinite leads are built out of Au(111) nanowires.

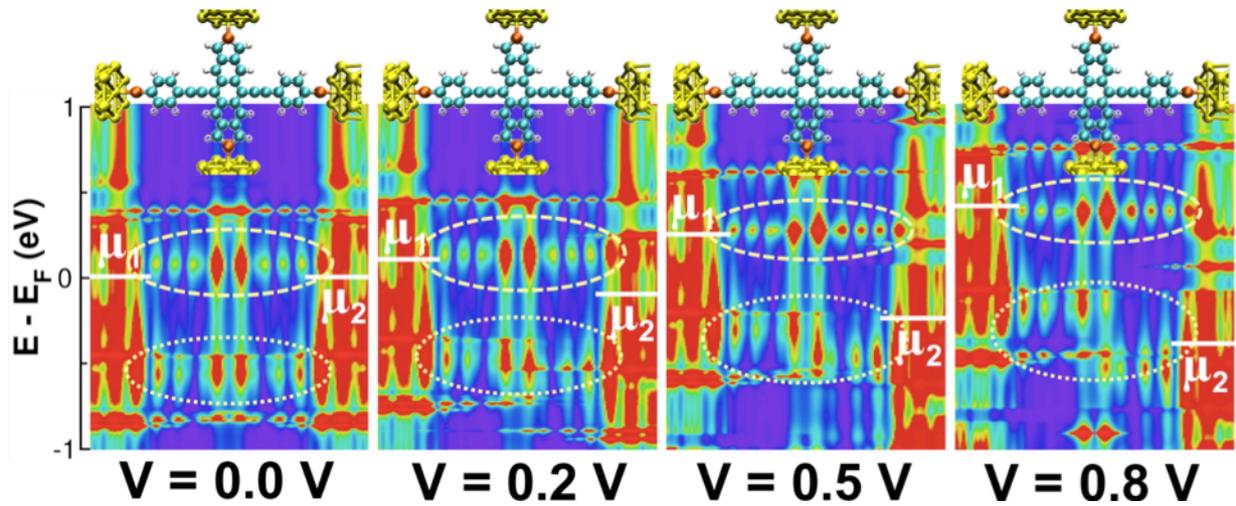

**FIG. 2:** (Color) Position-dependent density of states (DOS) along the left-to-right direction at biases of 0.0, 0.2, 0.5 and 0.8 V. The DOS is averaged in the plane perpendicular to the current. Note that the electronic states originating from the vertical leads are not included in the figure. The chemical potentials of the left and right leads are shown as white lines. The zero of energy is at $(\mu_1 + \mu_2)/2$. The dotted and dashed ovals enclose the pseudo-molecular states that originate from the highest occupied and lowest unoccupied molecular orbitals, respectively.

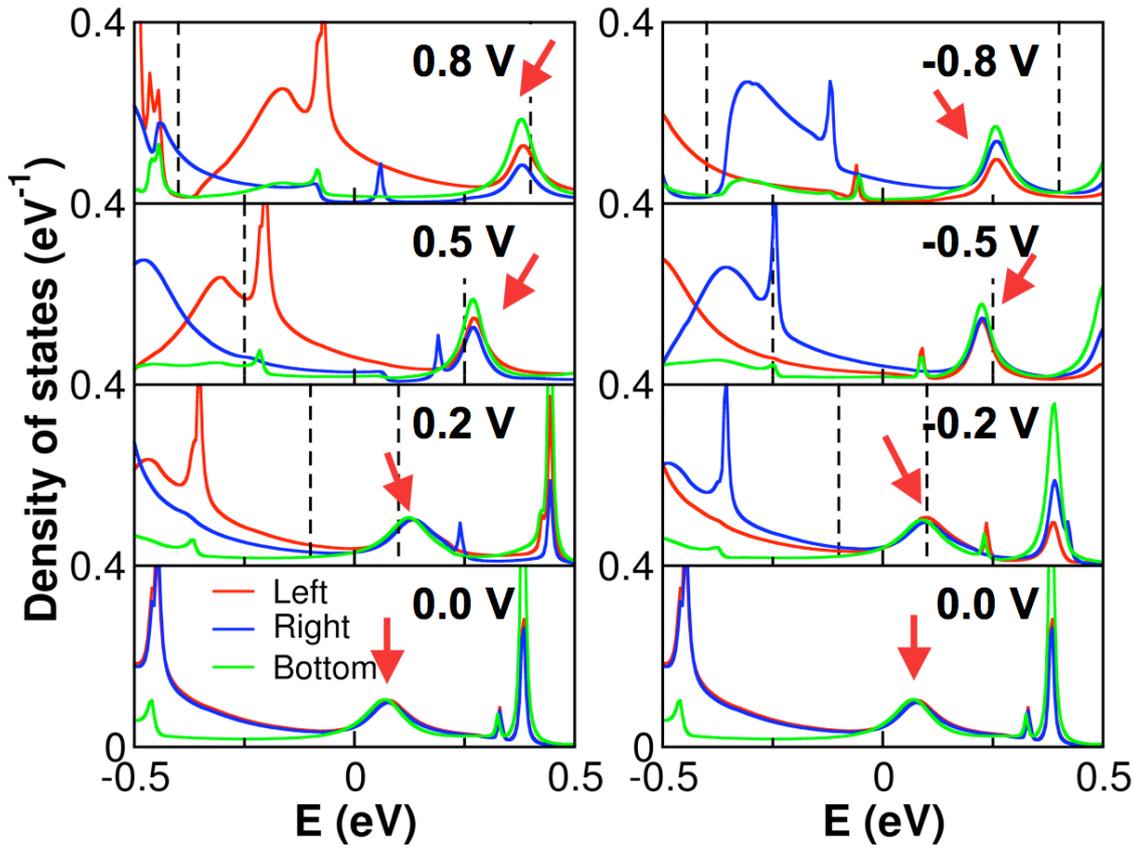

**FIG. 3:** (Color). Changes in the pseudo-molecular states with the bias voltage. The densities of states close to the left, right and bottom molecule-lead interfaces (of leads 1, 2, and 3) are shown. Due to the symmetry in the bias geometry, the electronic structures at the top and bottom interfaces are identical. The red arrow indicates the position of the LUMO, which mainly drives the current.

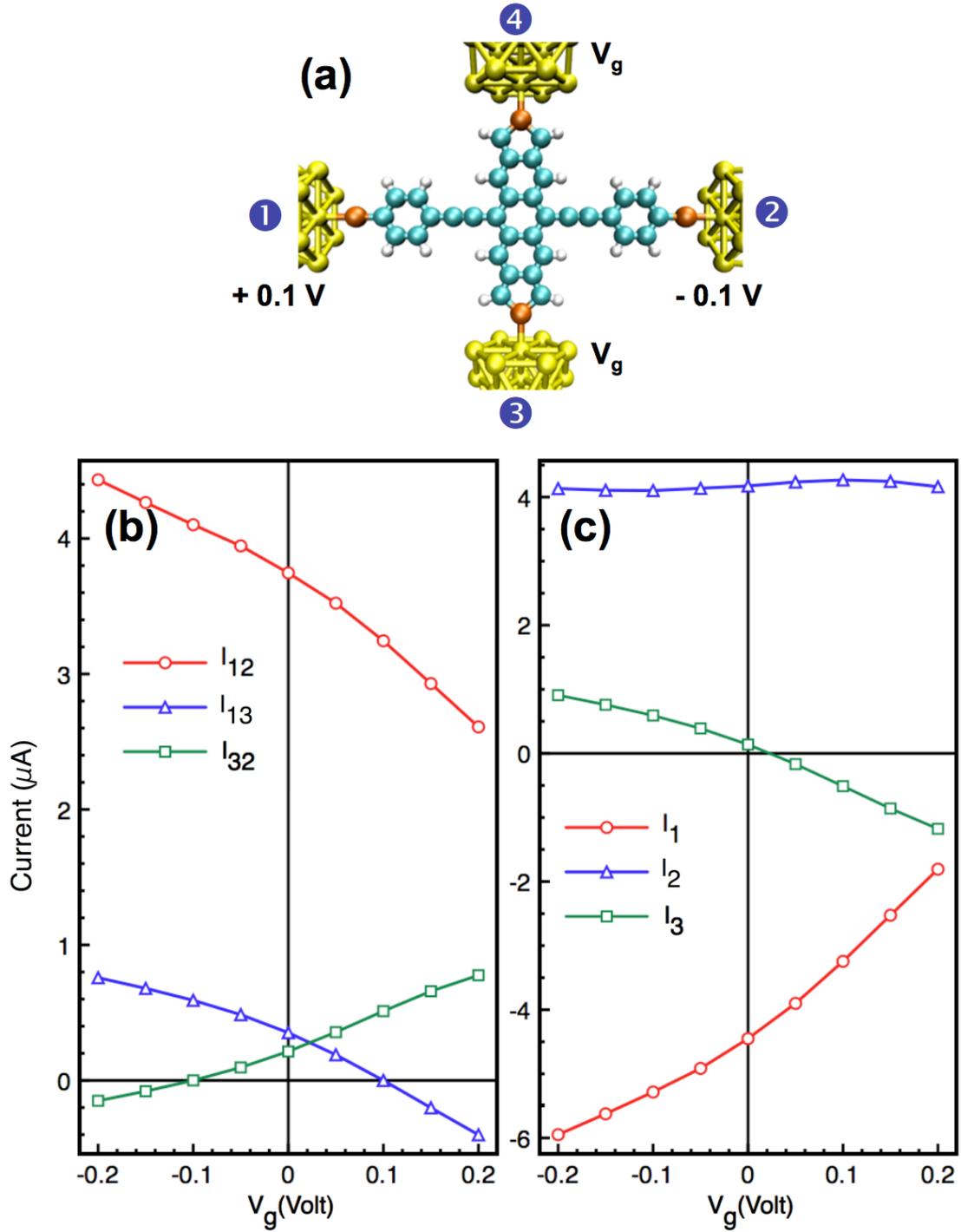

**FIG. 4:** (Color online) Currents in the four-terminal geometry shown in (a) as a function of pseudo-gate voltage $V_g$. (b) Currents $I_{ij}$ passing through terminals $i$ and $j$. A positive (negative) $I_{ij}$ indicates electrons flow from electrode $i$ ($j$) to electrode $j$ ($i$). (c) Total currents $I_i$ passing through terminals $i$. The currents $I_3$ and $I_4$ are equal by symmetry. $I_i > 0$ indicates electrons entering electrode $i$ and $I_i < 0$ indicates electrons leaving electrode $i$.